\documentclass[conference]{IEEEtran}
\IEEEoverridecommandlockouts
\usepackage{tikz}
\usetikzlibrary{external}
\usepackage{cite}
\usepackage{amsmath,amssymb,amsfonts}
\usepackage{algorithmic}
\usepackage{graphicx}
\usepackage{textcomp}
\usepackage{lipsum}
\usepackage{setspace}
\usepackage{multirow}
\usepackage{array}
\usepackage{graphics}
\usepackage{pgfplots}
\usepackage{ntheorem }
\usepackage{flushend}
 \newcommand{\rmg}{\mathrm{g}}
 \newcommand{\rmT}{\mathrm{T}}
 \newcommand{\rmH}{\mathrm{H}}
 \newcommand{\rmu}{\mathrm{u}}
 \newcommand{\rmc}{\mathrm{c}}
 \newcommand{\rme}{\mathrm{e}}
 \newcommand{\rmd}{\mathrm{d}}
 \newcommand{\rms}{\mathrm{s}}
 \newcommand{\rmb}{\mathrm{b}}

 \newcommand{\diag}{\mathrm{diag}}
 \newcommand{\rmD}{\mathrm{D}}
 \newcommand{\rmRF}{\mathrm{RF}}
 \allowdisplaybreaks
 \newtheorem*{remark}{Remark}
\begin{document}
\makeatletter
\setlength{\abovedisplayskip}{4pt}
\setlength{\belowdisplayskip}{4pt}
\newcommand{\linebreakand}{%
\end{@IEEEauthorhalign}
\hfill\mbox{}\par
\mbox{}\hfill\begin{@IEEEauthorhalign}
}
\def\ps@IEEEtitlepagestyle{
	\def\@oddfoot{\mycopyrightnotice}
	\def\@evenfoot{}
}
\def\mycopyrightnotice{
	{\footnotesize \copyright~2020 IEEE\hfill} 
	\gdef\mycopyrightnotice{}
}
\makeatother
\title{Performance Analysis for Autonomous Vehicle 5G-Assisted Positioning in GNSS-Challenged Environments\\
\thanks{Copyright (c) 2020 IEEE. Personal use of this material is permitted. However, permission to use this material for any other purposes must be obtained from the IEEE by sending a request to pubs-permissions@ieee.org. G. Seco-Granados was supported in part by the Spanish Ministry of Science, Innovations and Universities through projects TEC2017-89925-R and TEC2017-90808-REDT. Henk Wymeersch was supported by the Swedish Research Council under grant 2018-03701.}
}
\author{\IEEEauthorblockN{Zohair Abu-Shaban}
\IEEEauthorblockA{\textit{University of New South Wales}\\
Canberra, Australia\\
zohair.abushaban@unsw.edu.au}
\and
\IEEEauthorblockN{Gonzalo Seco-Granados}
\IEEEauthorblockA{\textit{Universitat Aut\`onoma de Barcelona}\\
Barcelona, Spain \\
gonzalo.seco@uab.cat}
\linebreakand
\IEEEauthorblockN{Craig R. Benson}
\IEEEauthorblockA{\textit{University of New South Wales}\\
	Canberra, Australia\\
c.benson@unsw.edu.au}
\and
\IEEEauthorblockN{Henk Wymeersch}
\IEEEauthorblockA{\textit{Chalmers University of Technology}\\
Gothenburg, Sweden \\
henkw@chalmers.se}
}

\maketitle

\begin{abstract}
Standalone Global Navigation Satellite Systems (GNSS) are known to provide a positioning accuracy of a few meters in open sky conditions. This accuracy can drop significantly when the line-of-sight (LOS) paths to some GNSS satellites are obstructed, e.g., in urban canyons or underground tunnels. To overcome this issue, the general approach is usually to augment GNSS systems with other dedicated subsystems to help cover the gaps arising from obscured LOS. Positioning in 5G has attracted some attention lately, mainly due to the possibility to provide cm-level accuracy using 5G signals and infrastructure, effectively imposing no additional cost. In this paper, we study the hybridization of GNSS and 5G positioning in terms of achievable position and velocity error bounds. We focus on scenarios where satellite visibility is constrained by the environment geometry, and where the GNSS and 5G positioning systems fail to perform individually or provide prohibitively large error.
\end{abstract}


\section{Introduction}
Standalone code-driven Global Navigation Satellite Systems (GNSS) positioning methods provide an accuracy of a few meters. When augmented with complementary systems that provide carrier-phase reference signals such as Real-Time Kinematic (RTK) and Precise Point Positioning (PPP), position accuracy goes down to cm-level \cite{GNSSBook_accuracy_2017}. However, such augmentation systems suffer from two main shortcomings \cite{GNSSBook_drawbacks_2017} i) The convergence time for a position fix tends to be long, ii) they are prone to intermittent reception where carrier-phase information may be lost, warranting a new position fix calculation. Therefore, we alternatively propose the soon-to-be ubiquitous 5G millimeter-wave (mmWave) communication technology \cite{Andrews2014, Rappaport2013,Akyildiz2016,Heath2016} to assist GNSS receivers to cover the gaps arising from intermittent GNSS reception and/or provide backup system that could take over the positioning task in GNSS-challenged scenarios, such as urban canyons, indoors, underground tunnels or malicious jamming attacks. 

Motivated by the ever-increasing applications requiring location-awareness, many recent studies investigated 5G positioning with mmWave transmission \cite{Taranto2014, Guerra2017, Zohair2017}. All these studies showed that 5G standalone systems can provide cm-level positioning accuracy. 5G employs antenna arrays at the base stations and user equipment, and hence, high-accuracy positioning can be performed with a single base station through the estimation of the range or pseudorange (PR) and the directions of arrival and departure (DOA, DOD) \cite{Guerra2017, Zohair2017, Rico2018}. MmWave channels are highly sparse and 5G positioning would be viable in environments with many reflections, in which GNSS would fail or have a very poor performance. With all the advantages of 5G, it is highly sensitive to synchronization accuracy and may not be fully available in non-urban areas. Moreover, GNSS may still be the favorable method of positioning in some scenarios, especially in open areas and highways. Therefore, positioning based on synergies of GNSS and 5G helps solve the shortcomings of the individual systems, while retaining and improving the strengths of each. As a use case, we propose this hybridization for systems of autonomous vehicles (AVs), which will be equipped with 5G transceivers, in any case, making integrating 5G with GNSS a natural solution.

\begin{figure}[!t]
	\centering
	\includegraphics{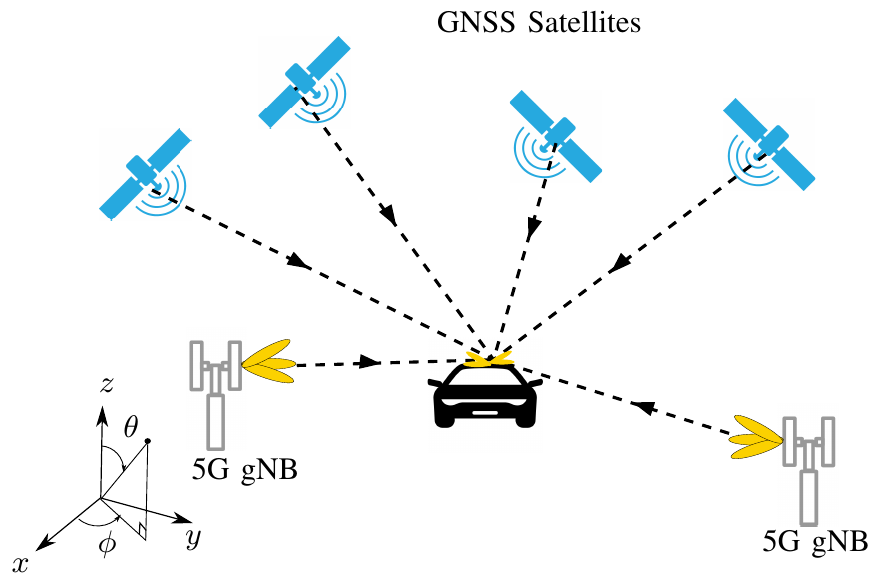}
	\caption{A diagram illustrating the concept of hybrid positioning using GNSS and 5G mmWave signals.}
	\label{fig:concept}
	\vspace{-5mm}
\end{figure}
In this paper, we focus on scenarios where the AV is exposed to line-of-sight (LOS) GNSS signals for a short duration, such as those in urban streets. We also consider scenarios where GNSS satellites are ill-positioned causing poor geometric dilution of precision (GDOP). In these cases, non-GNSS methods are more suited and 5G is considered in this paper based on the advantages discussed above. Fig. \ref{fig:concept} illustrates the concept of hybrid positioning using GNSS and 5G signals. We conduct a feasibility study on hybrid 5G-GNSS positioning systems through the derivation of position (PEB) and velocity error bounds (VEB). 

Towards that, we consider 5G mmWave transceivers with OFDM signaling. The 5G signal received at the AV is parameterized by the DOA, DOD, PR and Doppler shifts, while the signals received from GNSS satellites are parameterized by the PR and Doppler shifts. After deriving the Fisher information of the two sets of parameters, we obtain the Fisher information of position and velocity as a transformation of parameters based on geometrical relationships \cite{kay1993}. We then provide an analysis of when and how 5G can support GNSS, and what the parameters that govern the positioning performance are. We also provide insights and guidelines on how to design these parameters. Finally, We present a numerical investigation of the 5G performance in urban streets when GNSS has poor GDOP. Concretely, we set up a scenario comprising 5G base stations (gNBs) and GNSS satellites with a poor geometry due to the visibility constraints of the scenario. Ultimately, we show by simulation that the introduction of 5G positioning will boost the positioning availability and accuracy by a significant margin.

\section{System Model}
Consider an AV receiving downlink signals from $G$ gNBs located at $\mathbf{p}_g=[p_{g,x},p_{g,y},p_{g,z}]^\rmT, 1\leq g\leq{G}$, and $S$ GNSS satellites located at an initial position $\mathbf{p}_s=[p_{s,x},p_{s,y},p_{s,z}]^\rmT$, and moving with velocities $\mathbf{v}_s=[v_{s,x},v_{s,y},v_{s,z}]^\rmT$, $1\leq{s}\leq{S}$. The gNBs are assumed to be synchronized with the GNSS system but not with the AV, whose bias $b_\rmu$ is unknown. We assume that $\mathbf{p}_s$ and $\mathbf{v}_s$ are assumed fixed during the observation time. Denote the AV's initial position, velocity and azimuth rotation angle by $\mathbf{p}=[p_{x},p_{y},p_{z}]^\rmT, \mathbf{v}=[v_{x},v_{y},v_{z}]^\rmT$ and $\phi_0$, respectively. Without loss of generality, we take the position of the first gNB as the origin of the coordinate system. We consider a short observation window, over which the velocities of the AV and satellites are assumed to be constant. This is a reasonable assumption since vehicles generally move in speeds of up to 50 or 60 km/h ($\approx$ 13.9--16.6 m/s) in the considered scenario of urban streets.

\subsection{5G mmWave LOS OFDM Signal Model}
Consider $M$  OFDM symbols with duration $T_0$, including a cyclic prefix (CP) of duration $T_\mathrm{cp}$, sequentially transmitted over $N_\rmb$ beams with a carrier frequency $f_g$  and a subcarrier spacing $\Delta{f}$. The signal corresponding to the $m$-th OFDM symbol, $1\leq{m}\leq{M}$, received over the $k$-th subcarrier, $-K/2\leq{k}\leq{K/2}$, at the output of the receive beamforming is given in the frequency domain by
\begin{align}
\mathbf{y}_{k,m}=\sum_{g=1}^{G}&\mathbf{W}^{(g)^\rmH}_{k,m}\mathbf{H}^{(g)}_{k,m}\mathbf{F}^{(g)}_{k,m}\mathbf{z}^{(g)}_{k,m}\notag\\&+\mathbf{W}^{(g)^\rmH}_{k,m}\mathbf{n}^{(g)}_{k,m}\in{\mathbb{C}}^{N_\rms},\label{eq:5g_obs}
\end{align}
where $\mathbf{H}^{(g)}_{k,m}\in{\mathbb{C}}^{N_\rmu\times N_g}$ is the channel matrix, $\mathbf{F}^{(g)}_{k,m}=\mathbf{F}^{(g)}_\rmRF\mathbf{F}^{(g)}_{\rmD_{k,m}}\in{\mathbb{C}}^{N_g\times N_\rms},$ is the transmit beamforming matrix such that $\|\mathbf{F}^{(g)}_{k,m}\|_\mathrm{F}=1,\mathbf{F}^{(g)}_\rmRF\in{\mathbb{C}}^{N_g\times N_\rmb}$ and $\mathbf{F}_{\rmD_{k,m}}\in{\mathbb{C}}^{N_\rmb\times N_\rms}$ are the analog and digital transmit beamforming matrices, respectively. $\mathbf{z}^{(g)}_{k,m}\in{\mathbb{C}}^{N_\rms}$ is the vector of transmitted symbols, while  $N_\rmu$ is the number of antennas at the AV, $N_g$ is the number of antennas at the $g$-th gNB, $N_\rmb$ is the number of transmitted beams, and $N_\rms$ is the number of transmitted symbols. The matrix $\mathbf{W}^{(g)}_{k,m}=\mathbf{W}^{(g)}_\rmRF\mathbf{W}^{(g)}_{\rmD_{k,m}}\in{\mathbb{C}}^{N_\rmu\times N_\rms}$, such that $\|\mathbf{W}^{(g)}_{k,m}\|_\mathrm{F}=1,$ $\mathbf{W}^{(g)}_\rmRF\in{\mathbb{C}}^{N_\rmu\times N_\rmb}$ and $\mathbf{W}^{(g)}_{\rmD_{k,m}}\in{\mathbb{C}}^{N_\rmb\times N_\rms}$ are the analog and digital receive  beamforming matrices, respectively. The additive white Gaussian noise is denoted by $\mathbf{n}^{(g)}_{k,m}\sim\mathcal{CN}(0,N_0)\in{\mathbb{C}}^{N_\rmu}$, $N_0$ is the noise PSD. 

\begin{remark}
	 To simplify the notation, we drop the superscript $(g)$ from the model in \eqref{eq:5g_obs}. The signal corresponding to a specific gNB can be therefore obtained using the parameters related to that gNB. The subscript $\rmg$ is used to distinguish  the parameters related to gNBs from those related to the AV or GNSS satellites. This is similar to consider only one gNB, and treating different gNBs independently. 
\end{remark}

Based on this remark, the definition of the notation in \eqref{eq:5g_obs} is now explained. The channel matrix is given by
\begin{align}
\mathbf{H}_{k,m}=\kappa_{k,m}\mathbf{a}_{\rmu,k}(\theta_\rmu,\phi_\rmu)\mathbf{a}^\rmH_{\rmg,k}(\theta_\rmg,\phi_\rmg),
\end{align} 
where 
\begin{align}
\kappa_{k,m}\triangleq\sqrt{P_g N_g N_\rmu}\rme^{-j2\pi k\Delta{f}\tau_{\rmb_\rmg}}\rme^{j2\pi f_{\rmd_\rmg} T_0 m},
\end{align}
such that $|\kappa_{k,m}|^2=P_g N_g N_\rmu$, and $P_g$ is the average received power from the gNB. The complex channel gain is assumed to have been compensated for. $\tau_{\rmb_\rmg}$ and $f_{\rmd_\rmg}$ are \textit{biased} TOA \footnote{Consider a transmitted signal $\textbf{g}(t)$, where $t$ is taken with reference to the gNB's clock. The signal is received as $\textbf{g}(t-\tau+vt/c)$ where $\tau$ is the propagation delay and $v$ is the relative speed of motion. To write the received signal with reference to the AV's clock that is biased by $b_\rmu$ with respect to the gNB,  replace $t$ by $t-b_\rmu$ and use$f_{\rmd_\rmg}/f_g=v/c$ to obtain $\textbf{g}((1+f_{\rmd_\rmg}/f_g)(t-b_\rmu)-\tau)$. That is, the biased TOA $\tau_{\rmb_\rmg}=(1+f_{\rmd_\rmg}/f_g)b_\rmu+\tau\approx b_\rmu+\tau$.} and Doppler frequency. All the OFDM symbols are assumed to be delayed by the same TOA. $(\theta_\rmu,\phi_\rmu)$ and $(\theta_\rmg,\phi_\rmg)$ are the DOA and DOD, respectively.  $\mathbf{a}_{\rmu,k}(\theta_\rmu,\phi_\rmu)$ is the array response vectors of the AV defined by 
\begin{align}
\mathbf{a}_{\rmu,k}(\theta_\rmu,\phi_\rmu)=\frac{1}{\sqrt{N_\rmu}}\rme^{-j\frac{2\pi}{\lambda_k}\mathbf{L}_\rmu\mathbf{u}(\theta_\rmu,\phi_\rmu)}.
\end{align}
$\mathbf{L}_\rmu\in\mathbb{R}^{N_\rmu\times{3}}$ is the antenna location matrix in half-wavelength, $\lambda_k=\frac{c}{f_g+k\Delta f}$ and $\mathbf{u}(\theta,\phi)=[\cos\phi\sin\theta, \sin\phi\sin\theta, \cos\theta]^\mathrm{T}$ is a unit vector pointing towards an azimuth angle $\phi$ and an elevation angle $\theta$. $\mathbf{a}_{\rmg,k}(\theta_\rmg,\phi_\rmg)$ can be defined similarly. Note that in this model, the complex channel gain assumed to be estimated \textit{a priori}.

Finally, note that Doppler shift introduces a frequency error that may cause loss of sub-carrier orthogonality. Therefore, 
$\mathbf{z}_{k,m}\in{\mathbb{C}}^{N_\rms}$ is the signal transmitted on the $k$-th subcarrier including the interference from adjacent subcarriers and is modeled by  \cite{barbarossa2005}
\begin{align}
\mathbf{z}_{k,m}=\mathbf{x}_{k,m}+\sum_{k'\ne k}\mathbf{x}_{k',m}c_{k-k'}(f_{\rmd_\rmg} T_\rms)=\mathbf{X}_m\mathbf{c}_k,\label{eq:z}
\end{align}
where $\mathbf{X}_m\triangleq[\mathbf{x}_{-\frac{K}{2},m},\cdots,\mathbf{x}_{\frac{K}{2},m}]$ such that $\mathbf{x}_{k,m}\triangleq[X^{(1)}_{k,m},\cdots,X^{(N_\rms)}_{k,m}]^\rmT$ is the vector of $N_\rms$ transmitted symbols, with a duration $T_\rms$, and $\mathbf{c}_k$ is the $(k+1+K/2)$-th column of the circulant matrix $\mathbf{C}=j2\pi f_{\rmd_\rmg} T_s\mathbf{D}^\rmH\mathbf{Q}\mathbf{D}+\mathbf{I}_K,$ where $\mathbf{Q}=\diag(-K/2,-K/2+1,\cdots,K/2)$ and $\mathbf{D}$ is the DFT matrix \cite{barbarossa2005}.

\subsection{GNSS Satellite Signals}
Assuming that the Doppler frequency is much less than the carrier frequency\footnote{This assumption enables us write the signal using $t$ instead of $\left(1+\frac{f_{d_s}}{f_{s}}\right)t$}, the signal received from the $s$-th GNSS satellite, $1\leq{s}\leq{S}$, can be written as
\begin{align}
y_s(t)=\sqrt{P_\rms}x_s( t-\tau_{\rmb_s})\rme^{j2\pi f_{d_s}t}+n_s(t), \label{eq:GNSS_obs}
\end{align}
where $P_\rms, f_{\rmd_s}, \tau_{\rmb_s}$ and $\varphi_\rms$ are the received power, Doppler frequency and biased TOA and phase delay, respectively. $x_s(t)$ is the reference signal transmitted from the $s$-th satellite and modeled as
\begin{align}
x_s(t)=\sum_{\ell=0}^{N_\mathrm{so}-1}c_{s\ell}r\left(t-\ell T_\rmc\right),\label{eq:sat_tx_sig}
\end{align}
where $c_{s\ell}$ is the $\ell$-th PN-code chip with duration $T_c$, $r(t)$ is the pulse-shaping filter and $N_\mathrm{so}$ is the total number of transmitted chips. 
\section{PEB and VEB Derivation}
\subsection{Derivation of GNSS and 5G FIMs}
The vectors of unknowns associated with the $g$-th gNB signal and the $s$-th satellite can be written as
\begin{align}
\boldsymbol{\eta}_\rmg&\triangleq[\theta_\rmg,\phi_\rmg,\theta_\rmu,\phi_\rmu,\tau_{\rmb_\rmg},f_{\rmd_\rmg}]^\rmT\in\mathbb{R}^6,\label{eta_g}\\
\boldsymbol{\eta}_\rms&\triangleq[\tau_{\rmb_\rms},f_{\rmd_\rms}]^\rmT\in\mathbb{R}^2\label{eta_s}.
\end{align}
From \eqref{eq:5g_obs}, defining $\boldsymbol{\mu}_{k,m}\triangleq\mathbf{W}^\rmH_{k,m}\mathbf{H}_{k,m}\mathbf{F}_{k,m}\mathbf{z}_{k,m}$, then $\mathbf{J}_\rmg\in\mathbb{R}^{6\times{6}}$, the  Fisher information matrix of $\boldsymbol{\eta}_\rmg$, can be computed element-wise using \cite{kay1993}
\begin{align}
[\mathbf{J}_\rmg]_{a,b}=\frac{1}{N_0}\sum_{\forall k,m}\Re\left\{\frac{\partial\boldsymbol{\mu}^\rmH_{k,m}}{\partial\eta_{\overset{\mathrm{g}}{a}}}(\mathbf{W}_{k,m}^\rmH\mathbf{W}_{k,m})^{-1}\frac{\partial\boldsymbol{\mu}_{k,m}}{\partial\eta_{\overset{\mathrm{g}}{b}}}\right\}, \label{eq:FIM_5g}
\end{align}
where $\eta_{\overset{\mathrm{g}}{a}}$ is the $a$-th element in $\boldsymbol{\eta}_\rmg$, $1\leq a,b \leq 6$. The derivation of the elements of $\mathbf{J}_\rmg$ is  provided in Appendix \ref{derivation_FIM_5G}. Note how $\mathbf{z}_{k,m}$ depends on $f_{\rmd_\rmg}$, which is accounted for when the FIM is derived. See \eqref{z_f}.

Similarly,  defining ${\mu}_s(t)=\sqrt{P_\rms}x_s( t-\tau_{\rmb_s})\rme^{j2\pi f_{d_s}t}$ from \eqref{eq:GNSS_obs}, then $\mathbf{J}_s\in\mathbb{R}^{2\times{2}}$, the Fisher information matrix of $\boldsymbol{\eta}_\rms$ is given by
\begin{align}
[\mathbf{J}_s]_{a,b}=\frac{1}{N_0}\int_0^{T_\mathrm{so}}\Re\left\{\frac{\partial\mu_s^*(t)}{\partial\eta_{\overset{\mathrm{s}}{a}}}\frac{\partial\mu_s(t)}{\partial\eta_{\overset{\mathrm{g}}{b}}}\right\}\rmd t \label{eq:FIM_GNSS},
\end{align}
where $T_\mathrm{so}=N_\mathrm{so}T_\rmc$ is the satellite signal observation time and $1\leq a,b \leq 2$. Note that we assume observations from different satellites to be independent. As shown in Appendix \ref{app:fim_sat}
\begin{align}
\mathbf{J}_s=\frac{4\pi^2 P_\rms T_\mathrm{so} }{N_0}\begin{bmatrix}
W^2_\mathrm{eff}&0\\
0&T^2_\mathrm{eff}
\end{bmatrix},
\end{align}
where
\begin{align*}
W^2_\mathrm{eff}&\triangleq\frac{1}{T_\rmc}\int_{-W/2}^{W/2}f^2|R(f)|^2\rmd f,\\ T_\mathrm{eff}^2&\triangleq\int_0^{T_\rmc}\bar{t}^2|r(t)|^2\rmd t,
\end{align*}
$\bar{t}^2\triangleq\frac{1}{N_\mathrm{so}}\sum_{\ell=0}^{N_\mathrm{so}-1}(t+\ell T_\rmc)^2$ and $R(f)$ is the PSD of $r(t)$, assumed to be symmetric around $f=0$.

\subsection{Position and Velocity Error Bounds}
We are interested in the AV position $\mathbf{p}$, velocity $\mathbf{v}$ and clock bias $b_\rmu$, and consequently need to compute the FIM of
\begin{align}
\boldsymbol\eta'\triangleq[\mathbf{p}^\rmT,\mathbf{v}^\rmT, b_\rmu]^\rmT\in\mathbb{R}^7\label{eta_prime} 
\end{align} 
as a transformation of parameters. Given that $\mathbf{J}_\rmg$ and $\mathbf{J}_\rms$ provide independent information, they can be transformed separately as 
\begin{align}
\mathbf{J}_{\boldsymbol{\eta'}}=\underbrace{\sum_{g=1}^G\mathbf{T}_\rmg\mathbf{J}_{\rmg}\mathbf{T}_\rmg^\rmT}_\text{Information from 5G}+\underbrace{\sum_{s=1}^S\mathbf{T}_s\mathbf{J}_{s}\mathbf{T}_s^\rmT}_\text{Information from GNSS}\in\mathbb{R}^{7\times{7}},\label{tot_FIM}
\end{align}
where $\mathbf{T}_\rmg\triangleq\frac{\partial\boldsymbol{\eta}_\rmg^\rmT}{\partial\boldsymbol{\eta'}}
\in\mathbb{R}^{7\times{6}}$ and $\mathbf{T}_s\triangleq\frac{\partial\boldsymbol{\eta}_s^\rmT}{\partial\boldsymbol{\eta'}}
\in\mathbb{R}^{7\times{2}}$, obtained in Appendix \ref{App:C1} using the following formulas:
\begin{align*}
\theta_\rmg=&\cos^{-1}\left(\frac{p_z-p_{g,z}}{\|\mathbf{p}-\mathbf{p}_g\|}\right),\\
\phi_\rmg=&\tan^{-1}\left(\frac{p_y-p_{g,y}}{p_x-p_{g,x}}\right),\\
\theta_\rmu=&\cos^{-1}\left(\frac{-p_z+p_{g,z}}{\|\mathbf{p}-\mathbf{p}_g\|}\right),\\	
\phi_\rmu=&\tan^{-1}\left(\frac{p_y-p_{g,y}}{p_x-p_{g,x}}\right)-\phi_0-\pi\\
f_{\rmd_\rmg}=&-\frac{(\mathbf{v}-\mathbf{v}_g)^\rmT(\mathbf{p}-\mathbf{p}_g)}{\lambda_g\|\mathbf{p}-\mathbf{p}_\rmg\|},\\
\tau_{\rmb_\rmg}=&b_\rmu+\frac{\|\mathbf{p}-\mathbf{p}_\rmg\|}{c},\\
f_{\rmd_s}=&-\frac{(\mathbf{v}-\mathbf{v}_s)^\rmT(\mathbf{p}-\mathbf{p}_s)}{\lambda_s\|\mathbf{p}-\mathbf{p}_s\|},\\
\tau_{\rmb s} =&b_\rmu+\frac{\|\mathbf{p}-\mathbf{p}_s\|}{c}.
\end{align*} 
where $c=f_g\lambda_g=f_\rms\lambda_\rms$ is the speed of light. Note that for each gNB, $\mathbf{J}_{\rmg}$ and $\mathbf{T}_\rmg$ in \eqref{tot_FIM} are computed from \eqref{eq:FIM_5g} and \eqref{eq:trans_g}, respectively, using the parameters of that gNB.

Finally, we obtain $\mathbf{J}^\rme_{\mathbf{p,v}}$, the EFIM of $\mathbf{p}$ and $\mathbf{v}$,   by writing $\mathbf{J}_{\boldsymbol{\eta'}}$ in block form as
\begin{align*}
\mathbf{J}_{\boldsymbol{\eta'}}=
\begin{bmatrix}
\mathbf{J}_{\mathbf{p,v}}& \mathbf{J}_{\mathbf{pv},b_\rmu}\\
\mathbf{J}_{\mathbf{pv},b_\rmu}^\rmT&J_{b_\rmu}
\end{bmatrix}.
\end{align*}
Using Schur complement, we can derive
\begin{align*}
\mathbf{J}^\rme_{\mathbf{p,v}}=\mathbf{J}_{\mathbf{p,v}}-\frac{1}{J_{b_\rmu}}\mathbf{J}_{\mathbf{pv},b_\rmu}\mathbf{J}_{\mathbf{pv},b_\rmu}^\rmT\in\mathbb{R}^{6\times{6}}.
\end{align*}
Consequently, defining $\mathbf{c}=\mathrm{diag}\left\{\left(\mathbf{J}^\rme_{\mathbf{p,v}}\right)^{-1}\right\}$, then
\begin{subequations}
\begin{align}\label{eq:PEB_VEB_def}
\text{PEB}&\triangleq\sqrt{c_1+c_2+c_3},\\
\text{VEB}&\triangleq\sqrt{c_4+c_5+c_6}.
\end{align}
\end{subequations}

Note that for the position and velocity to be computed with no ambiguity, $\mathbf{J}_{\boldsymbol{\eta'}}$ must be rank 7. Since $\mathbf{J}_\rmg$ is rank 6, at least 2 gNBs are needed to obtain PEB and VEB based on 5G only, when the clock bias is unknown. Similarly, $\mathbf{J}_\rms$ is rank 2, which leads to the widely known fact that at least 4 satellites are needed to compute the position and velocity in GNSS positioning systems. In principle, hybridization allows us to use less than this number of satellites and gNBs, as we can use, for example, 1 gNB and 1 satellite. The satellite signal can be used to estimate the AV clock bias, while the gNB signals can be used to estimate the position and velocity. It is intuitive that incorporating more signals can boost the performance.
\section {Numerical Results}
\subsection{Geometry}\label{scenario_def}
Two scenarios related to different satellite arrangements are presented in this section: an open visibility scenario, and a constrained visibility scenario:
\begin{itemize}
	\item \textit{Scenario A: Open Visibility} This is a reference scenario where the AV receives LOS signals from 4 GNSS satellites observed at well-spaced locations. These locations are given in spherical coordinates (See Fig. \ref{fig:concept}) as $\mathbf{p}_s=(\rho,\theta,\phi)$, where $\rho=20.2\times{10}^6$ and $(\theta,\phi)= (35.2^\circ, 45^\circ), (35.2^\circ, -135^\circ), (57.3^\circ, 130^\circ)$ and $(57.37^\circ, -39.8^\circ)$, respectively.
	\item \textit{Scenario B: Constrained Visibility} In this visibility-constrained scenario, the AV receives LOS signals from 4 GNSS satellites that are almost aligned on an arc, that is, a narrow azimuth sector. Such a scenario can arise in central business districts and other suburban areas where high-rising buildings limit the duration and the visibility of LOS satellite links, causing positioning to be challenging. The 4 satellites are assumed to be located at $\mathbf{p}_s=(\rho,\theta,\phi)$, where $\rho=20.2\times{10}^6$ and $(\theta, \phi)= (45^\circ, 0.08^\circ), (5^\circ, -0.66^\circ), (17^\circ, 0.20^\circ)$ and $(25^\circ, -0.14^\circ)$, respectively.
\end{itemize}

The satellites are assumed to move at a speed of 3.9 km/s \cite{ESA}, of which a maximum of 1 km/s is in the radial direction. The tangential direction is chosen arbitrarily in the plane orthogonal  to the radial direction.

\begin{figure}[!t]
	\centering
	\includegraphics{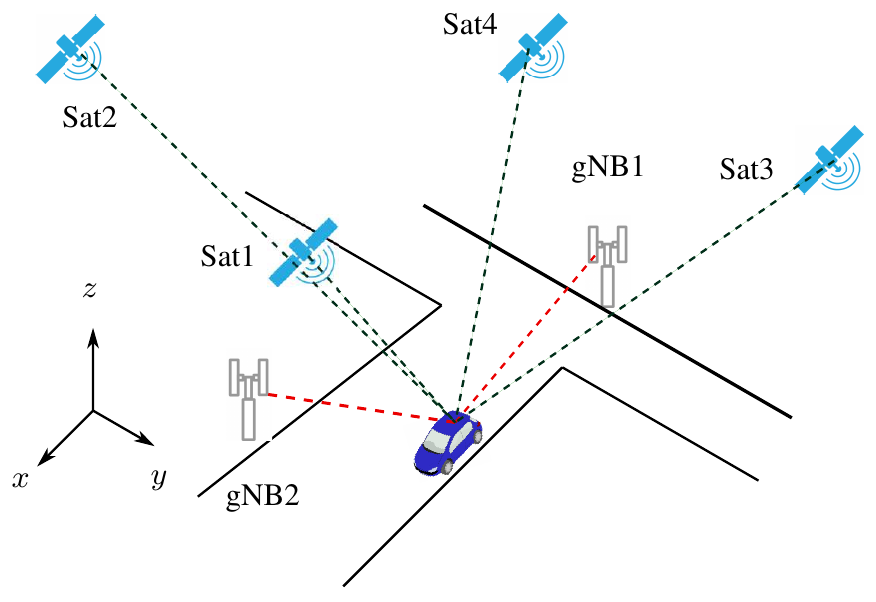}
	\caption{Scenario A: Four GNSS satellites observed at well-spaced locations.}
	\label{fig:scenario_A}
\end{figure}

\begin{figure}[!t]
	\centering
	\includegraphics{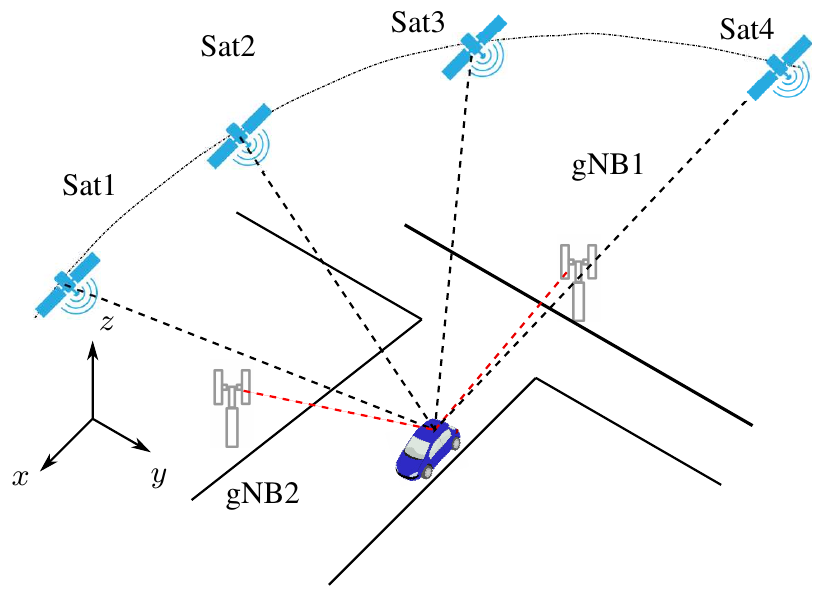}
	\caption{Scenario B: Four GNSS satellites with poor GDOP, resulting from the narrow azimuth sector observed.}
	\label{fig:ScenarioB}
\end{figure}
In both scenarios, we consider two gNBs, each equipped with an $12\times{12}$ uniform rectangular array, centered at $[0,0,7]^\rmT$ m  and $[20,-6,5]^\rmT$ m oriented towards the $+x$-axis and the $+y$-axis, respectively. Note that the origin of the coordinate systems is arbitrarily chosen to be on the ground under the first gNB.  The AV, equipped with an $8\times{8}$ array oriented toward the $+z$-axis, is assumed to be moving along the $x$-axis with a speed of $50$ km/h and measurements are taken when the AV is at $p_x=10$ m.
\subsection{Transceiver Parameters}
For the OFDM signals, we consider $K=1024$ subcarriers and $M=1000$ symbols transmitted over a carrier frequency 38 GHz and bandwidth of 125 MHz. That is, the observation time is 8.2 ms. The pilot samples are generated randomly as complex normal vectors such that $\|\textbf{x}_{m,k}\|^2=1$. The non-orthogonality of the subcarriers occurring due to the Doppler shift is assumed to affect one adjacent subcarrier on each side. That  is, the sum in \eqref{eq:z} is taken for $k'=k\pm1$. Finally, $P_g/N_0=30$ dBHz \cite{Pg_N0}.
\begin{table*}[!t]
	\caption{Observation time and bandwidth of GNSS and 5G systems.}
	\centering
	\begin{tabular}{c|c|c|c|c|}
		\cline{2-5}
		& \begin{tabular}[c]{@{}c@{}}Observation Time\\  (ms)\end{tabular} & \begin{tabular}[c]{@{}c@{}}Bandwidth\\ (MHz)\end{tabular}& \begin{tabular}[c]{@{}c@{}}$P/N_0$\\ (dBHz)\end{tabular} & \begin{tabular}[c]{@{}c@{}}Total Energy=\\Time$\times{}$Bandwidth$\times{}P/N_0$\end{tabular}\\ \hline
		\multicolumn{1}{|c|}{GNSS} &300&1.023& 40& 307$\times{10^{7}}$\\ \hline
		\multicolumn{1}{|c|}{5G}   &8.2 &125& 30&102.5$\times{10^{7}}$\\ \hline
	\end{tabular}
	\label{tab1}
\end{table*}

We consider L1 GNSS signals with a carrier frequency of 1575.42 MHz and a bandwidth of 1.023 MHz. The signals are received with a carrier-to-noise ratio of $P_s/N_0=$ 40 dBHz for a duration of 300 ms. The observation time and bandwidth of both systems are summarized in Table \ref{tab1}.

\subsection{Position Error Bounds}
Fig. \ref{fig:PEB_well} illustrates the PEB under various cases with well-positioned satellites as per Scenario A, defined in Section \ref{scenario_def}. With reference to the discussion below \eqref{eq:PEB_VEB_def}, hybrid positioning requires at least 1 gNB and 1 satellite, which is providing a PEB of 1.4 m in the studied scenario. 
The standalone GNSS PEB (4.25 m) represents the poorest case among those studied in Fig. \ref{fig:PEB_well}, but adding a single gNB brings PEB down to 75 cm. Adding a second gNB leads to further performance enhancement with a PEB of 2.5 cm.

The performance under Scenario B, whereby satellite locations cause poor GDOP is shown in Fig. \ref{fig:PEB_ill}. Not that with the exception of the standalone GNSS PEB,  Fig. \ref{fig:PEB_ill} shows that hybrid positioning in Scenario B provides a performance comparable to that under Scenario A. This implies that the performance of hybrid positioning is mainly governed by the abundant resources provided by 5G gNBS, as can be seen in Table \ref{tab1}.
\begin{figure}[!t]
	\centering
	\includegraphics{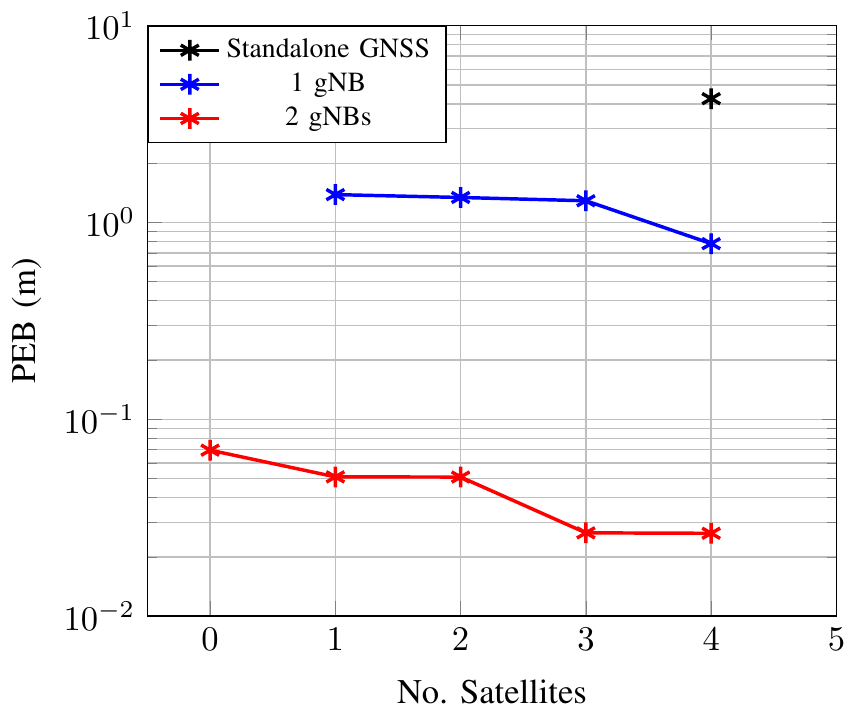}
	\caption{PEB under Scenario A comprising well-spaced GNSS satellites.}
	\label{fig:PEB_well}
\end{figure}

\begin{figure}[!t]
	\centering
	\includegraphics{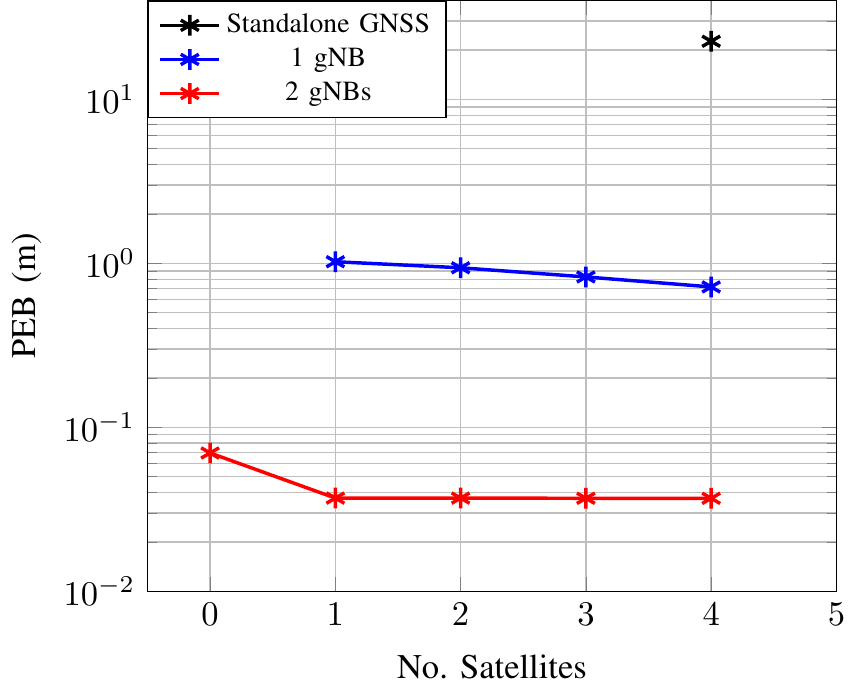}
	\caption{PEB under Scenario B comprising GNSS satellites with poor GDOP.}
	\label{fig:PEB_ill}
\end{figure}
\begin{figure}[!t]
	\centering
	\includegraphics{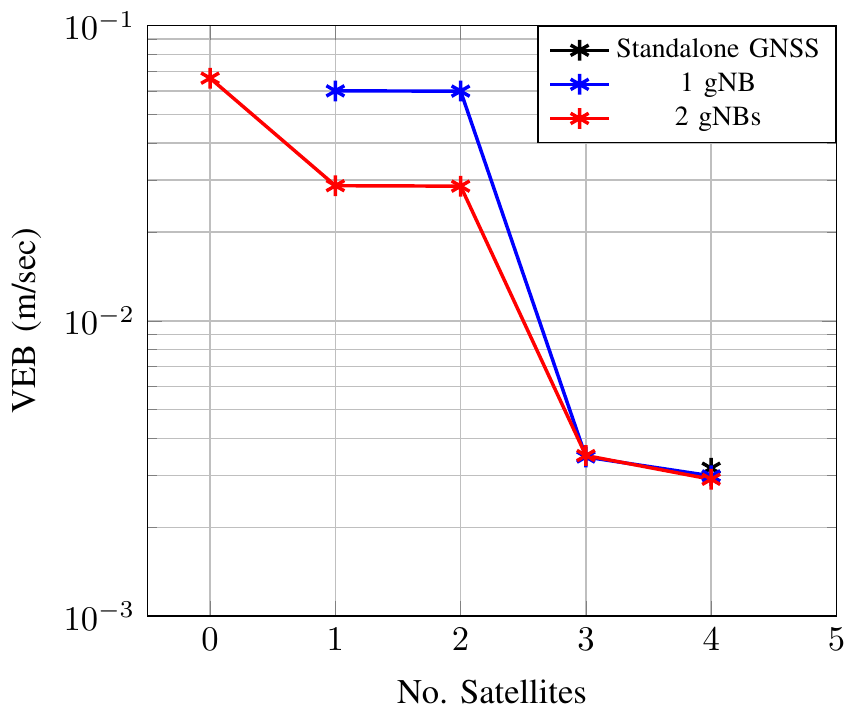}
	\caption{VEB under Scenario A comprising well-spaced GNSS satellites.}
	\label{fig:VEB_well}
\end{figure}
\subsection{Velocity Error Bounds}
From Fig. \ref{fig:VEB_well}, it can be seen that standalone GNSS system can provides a highly accurate velocity estimate, thanks to the long observation time and the good geometrical location of the satellites in Scenario A. However, although 5G provides a less accurate velocity estimate, it is in the order of a few centimeters, which is acceptable in systems of AVs.

Considering Fig. \ref{fig:VEB_ill}, it can be seen that when access to two gNBs or 4 satellites is not available, hybridization under ill-arranged satellites can be useful when access to only one gNB and 2-3 satellites is possible. Under the latter case, it is possible to obtain a VEB of  1.5--1.8 m/sec.

\begin{figure}[!t]
	\centering
	\includegraphics{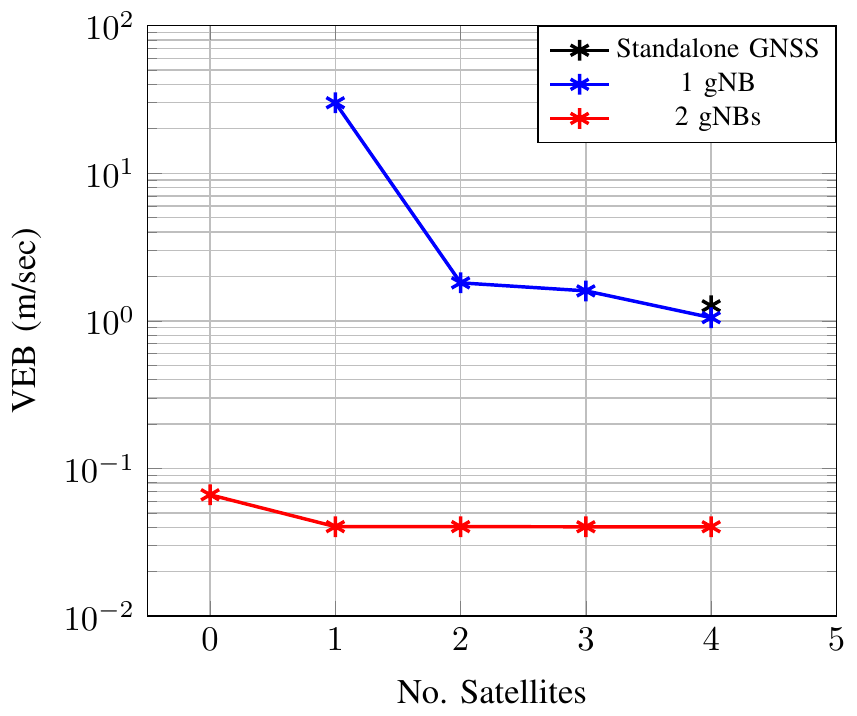}
	\caption{VEB under Scenario B comprising GNSS satellites with poor GDOP.}
	\label{fig:VEB_ill}
\end{figure}

\section{Conclusions}
This paper investigated the benefits of integrating 5G positioning with GNSS positioning. We presented theoretical results by deriving the position (PEB) and velocity (VEB) error bounds, and numerical results focusing on two possible scenarios whereby the satellites are well- and ill-positioned. Our numerical results show that hybridization of these two systems is beneficial when both  fail individually. it was also evident from our numerical results that when simultaneous access to 2 gNBs is available, precedence should be given to 5G positioning as it seems to provide satisfactory PEB and VEB. This can be attributed to the fact that the time-bandwidth resources available under 5G are more than triple of those available under GNSS (See Table \ref{tab1}). 

\appendices	

\section{Fisher Information Matrix of 5G Parameters}\label{derivation_FIM_5G}
We drop the angle parameters of $\mathbf{a}_{\rmu,k}$ and $\mathbf{a}_{\rmg,k}$ to simplify the notation. Then, defining 
\begin{align*}
\dot{\mathbf{a}}_{\rmg, \theta,k}\triangleq\frac{\partial\mathbf{a}_{\rmg,k}}{\partial\theta_\rmg},\qquad
\dot{\mathbf{a}}_{\rmg, \phi,k}\triangleq\frac{\partial\mathbf{a}_{\rmg,k}}{\partial\phi_\rmg},\\
\dot{\mathbf{a}}_{\rmu, \theta,k}\triangleq\frac{\partial\mathbf{a}_{\rmu,k}}{\partial\theta_\rmu},\qquad
\dot{\mathbf{a}}_{\rmu, \phi,k}\triangleq\frac{\partial\mathbf{a}_{\rmu,k}}{\partial\phi_\rmu},
\end{align*}
and noting that $\dot{\mathbf{z}}_{f_{\rmd_\rmg},k,m}\triangleq\frac{\partial\mathbf{z}_{\rmu,k}}{\partial f_{\rmd_\rmg}}=\mathbf{X}_m\dot{\mathbf{c}}_k,$ where $\dot{\mathbf{c}}_k$ is the $(k+\frac{K}{2}+1)$-th column of $j2\pi T_s\mathbf{D}^\rmH\mathbf{Q}\mathbf{D}$, we derive
\begin{subequations}\label{eq:Derivatives}
	\begin{align}
	\frac{\partial\boldsymbol{\mu}_{k,m}}{\partial\theta_\rmg}=&\kappa_{k,m}\mathbf{W}_{k,m}^\rmH\mathbf{a}_{\rmu,k}\dot{\mathbf{a}}_{\rmg, \theta,k}^\rmH\mathbf{F}_{k,m}\mathbf{z}_{k,m},\\
	\frac{\partial\boldsymbol{\mu}_{k,m}}{\partial\phi_\rmg}=&\kappa_{k,m}\mathbf{W}_{k,m}^\rmH\mathbf{a}_{\rmu,k}\dot{\mathbf{a}}_{\rmg, \phi,k}^\rmH\mathbf{F}_{k,m}\mathbf{z}_{k,m},\\
	\frac{\partial\boldsymbol{\mu}_{k,m}}{\partial\theta_\rmu}=&\kappa_{k,m}\mathbf{W}_{k,m}^\rmH\dot{\mathbf{a}}_{\rmu, \theta,k}\mathbf{a}_{\rmg,k}^\rmH\mathbf{F}_{k,m}\mathbf{z}_{k,m},\\
	\frac{\partial\boldsymbol{\mu}_{k,m}}{\partial\phi_\rmu}=&\kappa_{k,m}\mathbf{W}_{k,m}^\rmH\dot{\mathbf{a}}_{\rmu, \phi,k}\mathbf{a}_{\rmg,k}^\rmH\mathbf{F}_{k,m}\mathbf{z}_{k,m},\\
	\frac{\partial\boldsymbol{\mu}_{k,m}}{\partial\tau_{\rmb_\rmg}}=&(-j2\pi k\Delta{f})\kappa_{k,m}\mathbf{W}_{k,m}^\rmH\mathbf{a}_{\rmu,k}\mathbf{a}_{\rmg,k}^\rmH\mathbf{F}_{k,m}\mathbf{z}_{k,m},\\
	\frac{\partial\boldsymbol{\mu}_{k,m}}{\partial f_{\rmd_\rmg}}=&\kappa_{k,m}\mathbf{W}_{k,m}^\rmH\mathbf{a}_{\rmu,k}\mathbf{a}_{\rmg,k}^\rmH\mathbf{F}_{k,m}\times{}\notag\\&\left((j2\pi T_0 m)\mathbf{z}_{k,m}+\dot{\mathbf{z}}_{f_{\rmd_\rmg},k,m}\right).\label{z_f}
	\end{align}
\end{subequations}

For compactness, we also introduce the following notation 
\begin{align*}
\gamma_0&\triangleq\frac{P_g N_g N_\rmu}{N_0},\\
\alpha_f&\triangleq-j2\pi \Delta{f},\\
\breve{\mathbf{F}}_{k,m}&=\mathbf{F}_{k,m}\mathbf{z}_{k,m}\mathbf{z}_{k,m}^\rmH\mathbf{F}_{k,m}^\rmH,\\
\breve{\dot{\mathbf{F}}}_{k,m}&=\mathbf{F}_{k,m}\dot{\mathbf{z}}_{k,m}\mathbf{z}_{k,m}^\rmH\mathbf{F}_{k,m}^\rmH,\\
\breve{\ddot{\mathbf{F}}}_{k,m}&=\mathbf{F}_{k,m}\dot{\mathbf{z}}_{k,m}\dot{\mathbf{z}}_{k,m}^\rmH\mathbf{F}_{k,m}^\rmH,\\
\breve{\mathbf{W}}_{k,m}&=\mathbf{W}_{k,m}(\mathbf{W}_{k,m}^\rmH\mathbf{W}_{k,m})^{-1}\mathbf{W}_{k,m}^\rmH,\\
\dot{\mathbf{G}}_{k,m}&\triangleq j2\pi T_0m \breve{\mathbf{F}}_{k,m}+\breve{\dot{\mathbf{F}}}_{k,m},\\
 \ddot{\mathbf{G}}_{k,m}&\triangleq(2\pi T_0m)^2\breve{\mathbf{F}}_{k,m}-4\pi T_0 m\Im\{ \breve{\dot{\mathbf{F}}}_{k,m}\}+\breve{\ddot{\mathbf{F}}}_{k,m}.
\end{align*}
Next, the elements of \eqref{eq:FIM_5g} can be obtained using  \eqref{eq:Derivatives} as
\begin{subequations}
	\begin{align*}
	J_{\theta_\rmg}=&\gamma_0\sum_{\forall k,m}\Re\left\{(\dot{\mathbf{a}}_{\rmg, \theta,k}^\rmH\breve{\mathbf{F}}_{k,m}\dot{\mathbf{a}}_{\rmg, \theta,k})(\mathbf{a}_{\rmu,k}^\rmH\breve{\mathbf{W}}_{k,m}\mathbf{a}_{\rmu,k})\right\},\\
	J_{\theta_\rmg\phi_\rmg}=&\gamma_0\sum_{\forall k,m}\Re\left\{(\dot{\mathbf{a}}_{\rmg, \phi,k}^\rmH \breve{\mathbf{F}}_{k,m}\dot{\mathbf{a}}_{\rmg, \theta,k}^\rmH)( \mathbf{a}_{\rmu,k}^\rmH\breve{\mathbf{W}}_{k,m}\mathbf{a}_{\rmu,k})\right\},\\
	J_{\theta_\rmg\theta_\rmu}=&\gamma_0\sum_{\forall k,m}\Re\left\{(\mathbf{a}_{\rmg, k}^\rmH \breve{\mathbf{F}}_{k,m}\dot{\mathbf{a}}_{\rmg, \theta,k}^\rmH)(\mathbf{a}_{\rmu, k}^\rmH\breve{\mathbf{W}}_{k,m}\dot{\mathbf{a}}_{\rmu, \theta,k}^\rmH)\right\},\\
	J_{\theta_\rmg\phi_\rmu}=&\gamma_0\sum_{\forall k,m}\Re\left\{(\mathbf{a}_{\rmg, k}^\rmH \breve{\mathbf{F}}_{k,m}\dot{\mathbf{a}}_{\rmg, \theta,k}^\rmH)(\mathbf{a}_{\rmu, k}^\rmH\breve{\mathbf{W}}_{k,m}\dot{\mathbf{a}}_{\rmu, \phi,k}^\rmH)\right\},\\
	J_{\theta_\rmg\tau_{\rmb_\rmg}}=&\gamma_0\sum_{\forall k,m}\Re\left\{\alpha_f k(\mathbf{a}_{\rmg, k}^\rmH \breve{\mathbf{F}}_{k,m}\dot{\mathbf{a}}_{\rmg, \theta,k}^\rmH)( \mathbf{a}_{\rmu,k}^\rmH\breve{\mathbf{W}}_{k,m}\mathbf{a}_{\rmu,k})\right\},\\
	J_{\theta_\rmg f_{\rmd_\rmg}}=&\gamma_0\sum_{\forall k,m}\Re\left\{(\mathbf{a}_{\rmg, k}^\rmH\dot{\mathbf{G}}_{k,m}\dot{\mathbf{a}}_{\rmg, \theta,k})( \mathbf{a}_{\rmu,k}^\rmH\breve{\mathbf{W}}_{k,m}\mathbf{a}_{\rmu,k})\right\},\\
	J_{\phi_\rmg}=&\gamma_0\sum_{\forall k,m}\Re\left\{(\dot{\mathbf{a}}_{\rmg, \phi,k}^\rmH \breve{\mathbf{F}}_{k,m}\dot{\mathbf{a}}_{\rmg, \phi,k})( \mathbf{a}_{\rmu,k}^\rmH\breve{\mathbf{W}}_{k,m}\mathbf{a}_{\rmu,k})\right\},\\
	J_{\phi_\rmg\theta_\rmu}=&\gamma_0\sum_{\forall k,m}\Re\left\{(\mathbf{a}_{\rmg,k}^\rmH \breve{\mathbf{F}}_{k,m}\dot{\mathbf{a}}_{\rmg,\phi,k})(\mathbf{a}_{\rmu,k}^\rmH\breve{\mathbf{W}}_{k,m}\dot{\mathbf{a}}_{\rmu,\theta,k})\right\},\\
	J_{\phi_\rmg\phi_\rmu}=&\gamma_0\sum_{\forall k,m}\Re\left\{(\mathbf{a}_{\rmg,k}^\rmH \breve{\mathbf{F}}_{k,m}\dot{\mathbf{a}}_{\rmg,\phi,k})(\mathbf{a}_{\rmu,k}^\rmH\breve{\mathbf{W}}_{k,m}\dot{\mathbf{a}}_{\rmu,\phi,k})\right\},\\
	J_{\phi_\rmg\tau_{\rmb_\rmg}}=&\gamma_0\sum_{\forall k,m}\Re\left\{\alpha_f k(\mathbf{a}_{\rmg,k}^\rmH \breve{\mathbf{F}}_{k,m}\dot{\mathbf{a}}_{\rmg,\phi,k})( \mathbf{a}_{\rmu,k}^\rmH\breve{\mathbf{W}}_{k,m}\mathbf{a}_{\rmu,k})\right\},\\
	J_{\phi_\rmg f_{\rmd_\rmg}}=&\gamma_0\sum_{\forall k,m}\Re\left\{(\mathbf{a}_{\rmg, k}^\rmH\dot{\mathbf{G}}_{k,m}\dot{\mathbf{a}}_{\rmg, \phi,k})( \mathbf{a}_{\rmu,k}^\rmH\breve{\mathbf{W}}_{k,m}\mathbf{a}_{\rmu,k})\right\},\\
	J_{\theta_\rmu}=&\gamma_0\sum_{\forall k,m}\Re\left\{(\mathbf{a}_{\rmg,k}^\rmH \breve{\mathbf{F}}_{k,m}\mathbf{a}_{\rmg,k})( \dot{\mathbf{a}}_{\rmu, \theta,k}^\rmH\breve{\mathbf{W}}_{k,m}\dot{\mathbf{a}}_{\rmu, \theta,k})\right\},\\
	J_{\theta_\rmu\phi_\rmu}=&\gamma_0\sum_{\forall k,m}\Re\left\{(\mathbf{a}_{\rmg,k}^\rmH \breve{\mathbf{F}}_{k,m}\mathbf{a}_{\rmg,k})(\dot{\mathbf{a}}_{\rmu, \theta,k}^\rmH\breve{\mathbf{W}}_{k,m}\dot{\mathbf{a}}_{\rmu, \phi,k})\right\},\\
	J_{\theta_\rmu\tau_{\rmb_\rmg}}=&\gamma_0\sum_{\forall k,m}\Re\left\{\alpha_f k(\mathbf{a}_{\rmg,k}^\rmH \breve{\mathbf{F}}_{k,m}\mathbf{a}_{\rmg,k})(\dot{\mathbf{a}}_{\rmu, \theta,k}^\rmH\breve{\mathbf{W}}_{k,m}\mathbf{a}_{\rmu, k})\right\},\\
	J_{\theta_\rmu f_{\rmd_\rmg}}=&\gamma_0\sum_{\forall k,m}\Re\left\{(\mathbf{a}_{\rmg, k}^\rmH \dot{\mathbf{G}}_{k,m} \mathbf{a}_{\rmg,k})(\dot{\mathbf{a}}_{\rmu,\theta,k}^\rmH\breve{\mathbf{W}}_{k,m}\mathbf{a}_{\rmu,k})\right\},\\
	J_{\phi_\rmu}=&\gamma_0\sum_{\forall k,m}\Re\left\{(\mathbf{a}_{\rmg,k}^\rmH \breve{\mathbf{F}}_{k,m}\mathbf{a}_{\rmg,k})( \dot{\mathbf{a}}_{\rmu, \phi,k}^\rmH\breve{\mathbf{W}}_{k,m}\dot{\mathbf{a}}_{\rmu, \phi,k})\right\},\\
	J_{\phi_\rmu\tau_{\rmb_\rmg}}=&\gamma_0\sum_{\forall k,m}\Re\left\{\alpha_f k(\mathbf{a}_{\rmg,k}^\rmH \breve{\mathbf{F}}_{k,m}\mathbf{a}_{\rmg,k})(\dot{\mathbf{a}}_{\rmu, \phi,k}^\rmH\breve{\mathbf{W}}_{k,m}\mathbf{a}_{\rmu, k})\right\},\\
	J_{\phi_\rmu f_{\rmd_\rmg}}=&\gamma_0\sum_{\forall k,m}\Re\left\{(\mathbf{a}_{\rmg, k}^\rmH\dot{\mathbf{G}}_{k,m}\mathbf{a}_{\rmg,k})( \dot{\mathbf{a}}_{\rmu,\phi,k}^\rmH\breve{\mathbf{W}}_{k,m}\mathbf{a}_{\rmu,k})\right\},\\
	J_{\tau_{\rmb_\rmg}}=&\gamma_0|\alpha_f|^2\sum_{\forall k,m}\Re\left\{k^2(\mathbf{a}_{\rmg,k}^\rmH \breve{\mathbf{F}}_{k,m}\mathbf{a}_{\rmg,k})( \mathbf{a}_{\rmu,k}^\rmH\breve{\mathbf{W}}_{k,m}\mathbf{a}_{\rmu,k})\right\},\\
	J_{\tau_{\rmb_\rmg} f_{\rmd_\rmg}}=&-\gamma_0\sum_{\forall k,m}\Re\left\{\alpha_f k(\mathbf{a}_{\rmg, k}^\rmH\mathbf{a}_{\rmg,k})( \mathbf{a}_{\rmu,k}^\rmH\breve{\mathbf{W}}_{k,m}\mathbf{a}_{\rmu,k})\right\},\\
	J_{f_{\rmd_\rmg}}=&\gamma_0\sum_{\forall k,m}\Re\left\{(\mathbf{a}_{\rmg,k}^\rmH \ddot{\mathbf{G}}_{k,m} \mathbf{a}_{\rmg,k})( \mathbf{a}_{\rmu,k}^\rmH\breve{\mathbf{W}}_{k,m}\mathbf{a}_{\rmu,k})\right\}\label{eq:fim_fefe}.
	\end{align*}
\end{subequations}
\section{FIM of Satellite Parameters}\label{app:fim_sat}
Starting with ${\mu}_s(t)=\sqrt{P_\rms}x_s( t-\tau_{\rmb_s})\rme^{j2\pi f_{d_s}t}$ and noting that $\frac{\partial}{\partial\tau_{\rmb s}}x_s(t-\tau_{\rmb_s})=-\dot{x}_s( t-\tau_{\rmb_s}),$ then
\begin{align*}
\frac{\partial{\mu}_s(t)}{\partial\tau_{\rmb s}}&=-\sqrt{P_\rms}\dot{x}_s(t-\tau_{\rmb_s})\rme^{j2\pi f_{d_s}t},\\
\frac{\partial{\mu}_s(t)}{\partial f_{\rmd_\rmb s}}&=j2\pi t\sqrt{P_\rms}{x}_s(t-\tau_{\rmb_s})\rme^{j2\pi f_{d_s}t}.
\end{align*}
Consequently, from \eqref{eq:FIM_GNSS},
\begin{subequations}
	\begin{align}
	J_{\tau_{\rmb s}}&=\frac{P_\rms}{N_0}\int_0^{T_\mathrm{so}}|\dot{x}_s(t-\tau_{\rmb_s})|^2\rmd t\\
	&= \frac{P_\rms}{N_0}\int_{- W/2}^{W/2}(2\pi f)^2|X(f)|^2\rmd f,\label{eq:parseval}\\
	&= \frac{N_\mathrm{so} P_\rms}{N_0}\int_{- W/2}^{W/2}(2\pi f)^2|R(f)|^2\rmd f,\label{eq:psd_X_R}\\
	&=\frac{4\pi^2 P_\rms T_\mathrm{so} }{N_0}W_\mathrm{eff}^2,
	\end{align}
\end{subequations}
where 
\begin{align}
W_\mathrm{eff}^2\triangleq\frac{1}{T_\rmc}\int_{- W/2}^{W/2} f^2|R(f)|^2\rmd f.
\end{align}
\eqref{eq:parseval} follows from Parseval's theorem, while \eqref{eq:psd_X_R} follows from \eqref{eq:sat_tx_sig}. For a rectangular pulse shape, it can be shown that $W_\mathrm{eff}^2=W^2/(2\pi^2)$. 
Similarly,
	\begin{align}
	J_{f_{\rmd s}}&= \frac{4\pi^2 P_\rms}{N_0}\int_0^{T_\mathrm{so}} t^2|x_\rms(t-\tau_{\rmb_s})|^2\rmd t\label{eq:parseval2}.
	\end{align}
By the expansion of the summation in \eqref{eq:sat_tx_sig}, and a change of variables, it can be shown that
\begin{align}
	J_{f_{\rmd s}}&=\frac{4\pi^2P_\rms T_\mathrm{so} T_\mathrm{eff}^2}{N_0},\label{eq:psd_X_R2}
\end{align}
where 
\begin{align}
T_\mathrm{eff}^2\triangleq\int_0^{T_\rmc}\left\{\frac{1}{N_\mathrm{so}}\sum_{\ell=0}^{N_\mathrm{so}-1}(t+\ell T_\rmc)^2\right\}|r(t)|^2\rmd t
\end{align}For a rectangular pulse, $T_\mathrm{eff}^2=T_\mathrm{so}^2/12$.

Note that $J_{\tau_{\rmb s}f_{\rmd_s}}=0,$ because the integrand below is imaginary
\begin{align}
J_{\tau_{\rmb s}f_{\rmd_s}}&=\frac{-2\pi P_\rms}{N_0}\int_0^{T_\mathrm{so}}\Re\left\{j t|{x}_s(t-\tau_{\rmb_s})|^2\right\}\rmd t=0.
\end{align}

\section{Non-Zero Entries of Transformation Matrices }\label{App:C1}
\subsection{5G Parameter Transformation Matrix $\mathbf{T}_\rmg$} Defining $\bar{\mathbf{p}}_g\triangleq\mathbf{p}-\mathbf{p}_g=[\bar{p}_{g,x},\bar{p}_{g,y},\bar{p}_{g,z}]^\rmT$ and $\bar{\mathbf{v}}_g\triangleq\mathbf{v}-\mathbf{v}_g=[\bar{v}_{g,x},\bar{v}_{g,y},\bar{v}_{g,z}]^\rmT$, then the non-zero elements in the transformation matrix $\mathbf{T}_\rmg$ can be shown to be
\begin{subequations}\label{eq:trans_g}
	\begin{align}
	\frac{\partial\theta_\rmg}{\partial\mathbf{p}}=&\frac{[\bar{p}_x\bar{p}_z\quad \bar{p}_y\bar{p}_z\quad -(\bar{p}_x+\bar{p}_y)^2]^\rmT}{\|\bar{\mathbf{p}}_g\|^2\sqrt{\bar{p}_x^2+\bar{p}_y^2}},\\
	\frac{\partial\phi_\rmg}{\partial\mathbf{p}}=&\frac{1}{\bar{p}_x^2+\bar{p}_y^2}[-\bar{p}_y\quad \bar{p}_x\quad 0]^\rmT,\\
	\frac{\partial\theta_\rmu}{\partial\mathbf{p}}=&-\frac{\partial\theta_\rmg}{\partial\bar{\mathbf{p}}_g},\\
	\frac{\partial\phi_\rmu}{\partial\mathbf{p}}=&\frac{\partial\phi_\rmg}{\partial\bar{\mathbf{p}}_g},\\
	\frac{\partial f_{\rmd_\rmg}}{\partial\mathbf{p}}=&\frac{(\bar{\mathbf{v}}_g^\rmT\bar{\mathbf{p}}_g)\bar{\mathbf{p}}_g-\|\bar{\mathbf{p}}_g\|^2\bar{\mathbf{v}}_g}{\lambda_g\|\bar{\mathbf{p}}_g\|^3},\\
	\frac{\partial\tau_{\rmb_\rmg}}{\partial\mathbf{p}}=&\frac{\bar{\mathbf{p}}_g}{c\|\mathbf{p}\|}+\frac{b_\rmu}{f_g}\frac{\partial f_{\rmd_\rmg}}{\partial\mathbf{p}}\\
	\frac{\partial f_{\rmd_\rmg}}{\partial\mathbf{v}}=&-\frac{\bar{\mathbf{p}}_g}{\lambda_g\|\mathbf{p}\|},\\
	\frac{\partial\tau_{\rmb_\rmg}}{\partial\mathbf{v}}=&\frac{b_\rmu}{f_g}\frac{\partial f_{\rmd_\rmg}}{\partial\mathbf{v}}\\
	\frac{\partial\tau_{\rmb_\rmg}}{\partial b_\rmu}=&1+\frac{f_{\rmd_\rmg}}{f_g}.
	\end{align}
\end{subequations}
\subsection{5G Parameter Transformation Matrix $\mathbf{T}_\rmg$} Defining 
$\bar{\mathbf{p}}_s\triangleq\mathbf{p}-\mathbf{p}_\rms$ and $\bar{\mathbf{v}}_s\triangleq\mathbf{v}-\mathbf{v}_\rms$, then the non-zero elements in the transformation matrix $\mathbf{T}_s$ can be shown to be
\begin{subequations}\label{eq:trans_gnss}
	\begin{align}
	\frac{\partial f_{\rmd_s}}{\partial\mathbf{p}}=&\frac{(\bar{\mathbf{v}}_s^\rmT\bar{\mathbf{p}}_s)\bar{\mathbf{p}}_s-\|\bar{\mathbf{p}}_s\|^2\bar{\mathbf{v}}_s}{\lambda_s\|\bar{\mathbf{p}}_s\|^3},\\
	\frac{\partial\tau_{\rmb_\rms}}{\partial\mathbf{p}}=&\frac{\bar{\mathbf{p}}_s}{c\|\bar{\mathbf{p}}_s\|}+\frac{b_\rmu}{f_s}\frac{\partial f_{\rmd_s}}{\partial\mathbf{p}}\\
	\frac{\partial f_{\rmd_s}}{\partial\mathbf{v}}=&-\frac{\bar{\mathbf{p}}_s}{\lambda_s\|\bar{\mathbf{p}}_s\|},\\
	\frac{\partial\tau_{\rmb_\rms}}{\partial\mathbf{v}}=&\frac{b_\rmu}{f_\rms}\frac{\partial f_{\rmd_s}}{\partial\mathbf{v}}\\
	\frac{\partial\tau_{\rmb_\rms}}{\partial  b_\rmu}=&1+\frac{f_{\rmd_s}}{f_s}.
	\end{align}
\end{subequations}
\bibliographystyle{IEEEtran}

\end{document}